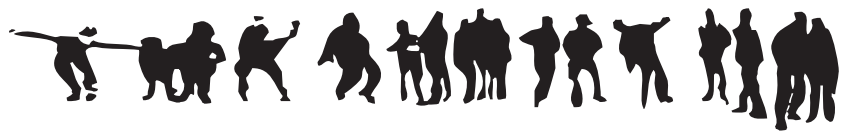

# L'acceptation et l'appropriation des ENT (Espaces Numériques de Travail) par les enseignants du primaire

▶**Elena CODREANU** (GREPS, Université Lyon2 ; LIRIS, INSA-Lyon ; Open Digital Education), **Christine MICHEL** (LIRIS, INSA-Lyon), **Marc-Eric BOBILLIER-CHAUMON** (GREPS, Université Lyon2), **Olivier VIGNEAU** (Open Digital Education)

■**RÉSUMÉ** • Cet article présente une évaluation des conditions d'usage d'un ENT (appelé ONE) par des enseignants du primaire. Elle est réalisée à travers deux études utilisant la théorie de l'activité comme cadre d'analyse. La première évalue les activités réelles effectuées par les enseignants sur l'ENT en analysant thématiquement le contenu des publications faites sur l'ENT, afin de comprendre comment les utilisateurs s'approprient l'outil. La seconde évalue les facteurs d'acceptation et de refus de la technologie, en décrivant la manière dont les enseignants vivent et perçoivent le rôle que l'ENT peut jouer sur l'évolution de leurs pratiques professionnelles (en les maintenant, les transformant ou les restreignant). Ces études ont permis de montrer que l'appropriation de ONE s'est produite notamment à travers les activités pédagogiques et de communication avec les parents d'élèves. L'acceptation favorable de cet ENT est due à la facilité d'usage et l'adéquation de l'interface aux enseignants et jeunes enfants.

■**MOTS-CLÉS** • ENT, Pratiques, Usages, Enseignement primaire

■**ABSTRACT** • *This article presents an evaluation of the conditions of use of a VWE (Virtual Work Environment) by primary school teachers. To this end, we conducted two studies and used activity theory as theoretical framework. Our first study aims to assess real practices carried out with the VWE and analyzed publications'content in order to understand how users appropriate the tool. The second study describes how teachers perceive the role of the VWE in the evolution of their working practices (maintaining, transforming and restricting the existent practices). These studies indicate that technological*

**Elena CODREANU, Christine MICHEL, Marc-Eric BOBILLIER-CHAUMON, Olivier VIGNEAU**

*appropriation is achieved through instructional and communicational uses. The acceptance of this VWE is due to its ease of use and interface adequacy to teachers and young children.*

■**KEYWORDS** • *Virtual Work Environment, Practices, Uses, Primary Education*

## 1. Introduction

Un ENT est un « dispositif global fournissant à son usager un espace dédié à son activité dans le système éducatif. Il est un point d'entrée unifié pour accéder au système d'information pédagogique de l'école ou de l'établissement » (Ministère de l'Education Nationale, 2012). Il vise à favoriser d'une part la communication et les pratiques collaboratives entre les membres d'une communauté scolaire via des services comme le blog ou la messagerie, et d'autre part les pratiques pédagogiques via des services comme le cahier de texte numérique, le cahier multimédia ou le blog.

Ces ENT sont principalement déployés dans l'enseignement supérieur et secondaire. Ils commencent à être utilisés pour l'enseignement primaire. L'ambition de cet article est de comprendre la façon dont les enseignants du primaire intègrent l'ENT à leurs pratiques professionnelles : comment et par quels moyens se déploie cette intégration ? Les enseignants développent-ils des usages spécifiques et innovants à l'aide du dispositif ? De (nouvelles) pratiques pédagogiques ont-elles émergé ou été modifiées par l'usage de l'ENT ? Quelles difficultés et contraintes suscite-t-il dans l'exercice de l'activité de ces enseignants ? Cette étude se propose de décrire les modalités d'appropriation et d'acceptation de l'ENT. L'appropriation a été étudiée sur la base d'une analyse thématique des publications faites sur l'ENT de manière à déterminer les types d'activités et d'usages réalisés. L'acceptation a été abordée sur la base d'une série d'entretiens semi-directifs. Il s'agissait d'identifier les facteurs favorisant ou s'opposant à l'adoption de l'ENT chez les enseignants. Notre objectif est de faire des propositions d'amélioration sur la conception des ENT ainsi que sur la manière d'accompagner le développement des pratiques innovantes avec les ENT. De plus, dans la mesure où les ENT sont encore peu répandus dans l'enseignement primaire, il s'agit aussi de décrire les processus d'acceptation et d'appropriation à l'œuvre dans ce contexte scolaire particulier.

L'analyse du contexte éducationnel actuel montre que le métier de professeur des écoles suppose, en plus des tâches éducatives proprement dites, un travail important de communication des informations scolaires auprès des parents. Le parent est décrit comme un collaborateur au processus éducatif de l'enfant par les Inspections Générales de l'Éducation Nationale (IGEN, 2006). L'article L. 313-2 du Code de l'Éducation prévoit que « des relations d'information mutuelles sont établies entre les enseignants et chacune des familles des élèves. Elles ont notamment pour objet

**Elena CODREANU, Christine MICHEL, Marc-Eric BOBILLIER-CHAUMON, Olivier VIGNEAU**

de permettre à chaque famille, d'avoir connaissance des éléments d'appréciation de l'élève » (Code de l'Education, 2016). Ces tâches informationnelles ne sont pas totalement intégrées par les enseignants (IGEN, 2006) et souvent ces relations de coopération sont tensionnelles (Warzee *et al.,* 2006), (Bruillard, 2011). Les enseignants considèrent en effet souvent les parents comme illégitimes sur le plan de la pédagogie, et pas assez compétents pour les accompagner dans leurs tâches éducatives (Bruillard, 2011). Ce problème de communication entre l'école et les familles existait dans le système d'enseignement primaire indépendamment de l'apparition des ENT.

Des études sur le second degré (collèges et lycées) ont montré que certains enseignants n'ont que partiellement réussi à intégrer les ENT dans leurs pratiques professionnelles. Ainsi, Prieur et Steck (Prieur et Steck, 2011) indiquent qu'en dépit du fait que les enseignants reconnaissent l'utilité pédagogique des ENT, ils ne sont pas prêts à les adopter pour des raisons de manque d'ergonomie, d'absence de formation, de maîtrise insuffisance des outils informatiques, de surcharge de travail et d'une résistance à étendre « l'espace-temps scolaire » hors de l'école. Poyet et Genevois (Poyet et Genevois, 2010) identifient de leur côté des différences de culture : les ENT sont généralement pensés comme des outils de gestion pour les entreprises et ils manqueraient dès lors de « traduction » et de sens vis-à-vis du cadre scolaire. Un des moyens possibles est l'utilisation de métaphores scolaires (cahier de textes, casier) au lieu de termes bureautiques (messagerie, agenda). Poyet et Genevois montrent que la méconnaissance de l'outil et le manque de compréhension de l'utilité ou de l'intérêt pédagogique entraînent des phases d'expérimentation insatisfaisantes où les enseignants testent les différentes fonctionnalités, sans avoir toujours une représentation de leurs potentialités et de leurs limites spécifiques et conduisent à préférer l'abandon au profit, là encore, d'outils personnels connus (comme le mail personnel). Bruillard et Hourbette (Bruillard et Hourbette, 2008) évoquent la complexité du déploiement des ENT qui se passe au croisement de multiples acteurs : enseignants, parents, élèves, académies, collectivités locales, éditeurs de logiciels, ministère. Ils remarquent un paradoxe entre la volonté du ministère d'ouvrir l'école aux parents d'élèves, la faible implication réelle des parents et la crainte par les enseignants d'une présupposée intrusion des parents dans leurs choix pédagogiques. À ces difficultés s'ajoute l'absence de reconnaissance institutionnelle pour les enseignants qui utilisent l'ENT, une déresponsabilisation des acteurs du terrain suite à l'appel aux entreprises extérieures en charge de concevoir l'ENT ou le risque d'inégalité voir d'exclusion de certains parents, moins équipés et formés au numérique. Missonier (Missonier, 2008) explique ce constat à partir des processus de conception et de déploiement des projets ENT qui sont pris en charge par les collectivités locales et les prestataires. Ces démarches manquent souvent d'efficacité car la résolution des controverses liées aux fonctionnalités ou aux usages est limitée par une transparence insuffisante du chef de projet. Cela finit par conduire à la diminution de l'engagement des différents acteurs du réseau. Prieur et Steck (Prieur et Steck, 2011) recommandent d'ouvrir des espaces de réflexions « articulant les pratiques actuelles des enseignants, les pratiques favorables pour accompagner l'apprentissage des compétences et les potentialités des différents outils de l'ENT de façon à construire des instrumentalisations possibles » ; l'objectif étant de trouver comment mieux adapter les usages prescrits aux contextes particuliers d'enseignement.

Voulgre (Voulgre, 2011) introduit une dimension politique. Les enseignants adhèrent a priori aux arguments favorables à l'utilisation des ENT : ils apparaissent ainsi utiles pour rattraper les cours (maladie, perte des notes), pour retrouver du travail antérieur ou accompagner des élèves en difficulté scolaire. Mais le fait que tous les enfants n'aient pas Internet à la maison représente une inégalité qui les freine dans l'usage et ce refus est vécu comme une « forme de contrepouvoir » envers les injonctions politiques. À l'inverse, l'un des facteurs d'acceptation est le respect de la hiérarchie, de l'institution et de la loi (obligation d'utiliser un ENT) ; d'autres concernent les valeurs de solidarité et d'entraide véhiculées par l'outil.

Louessard et Cottier (Louessard et Cottier, 2015) ont mis en évidence dans le secondaire les effets de la classe et des établissements sur les usages globaux des ENT. En effet, le choix des services ENT est différent d'une classe à une autre et d'un établissement à un autre et cela peut jouer un rôle important sur les usages des enseignants, des élèves et des parents d'élèves. Par exemple, le choix de certains enseignants de publier des informations sportives ou relatives aux voyages scolaires peut inciter les parents d'élèves à consulter l'ENT pour accéder à ces informations, malgré un certain refus initial à utiliser la plateforme. Ainsi, l'engagement et l'investissement des enseignants peuvent modifier la construction des pratiques autour de l'ENT.

Puimatto (Puimatto, 2009) dénonce une certaine « bureaucratisation » des ENT, qui sont principalement utilisés pour la gestion de la vie scolaire (notes et absences) et se prêteraient moins aux usages pédagogiques, auprès des élèves. Une possible explication serait la trop grande concentration de l'État sur l'industrialisation du projet et une vision macro-organisatrice, qui fait que les utilisateurs finaux et leurs besoins réels sont « oubliés ».

Schneewele (Schneewele, 2012) montre que les principaux bénéficiaires des ENT sont les parents d'élèves qui semblaient avoir très bien accepté l'outil et ses fonctionnalités, alors que les enseignants ont des opinions plus mitigées : ils semblent l'accepter plutôt suite aux injonctions hiérarchiques. Les usages des ENT seraient donc plus importants dans les collèges où l'ENT est tout simplement imposé par la direction. Les enseignants s'approprient généralement l'ENT à des fins de transmissions de documents ou d'informations alors que les parents et les élèves le font à des fins de consultation de documents. L'ENT serait donc un moyen incontournable pour se tenir informé. Les enseignants qui s'opposent à l'utilisation de l'ENT dénoncent une trop grande transparence de cet outil qui permettrait un contrôle et une mise en question de leurs choix pédagogiques par les inspecteurs scolaires ou par les parents d'élèves. Globalement, les usages (de nature communicationnelle ou pédagogique) restent en dessous des attentes des autorités. Il serait difficile de mettre en place de nouvelles pratiques qui incluent l'utilisation de l'ENT.

Compte tenu de ces obstacles rencontrés dans l'enseignement secondaire, nous souhaitons connaître et analyser les types d'usages faits dans le premier degré. Nous savons que les ENT sont des outils qui visent à réaliser des usages pédagogiques ou communicationnels (Ministère de l'Education Nationale, 2012). Nous avons vu cependant que les tâches communicationnelles ne sont pas totalement intégrées par les parents. Dans la mesure où les ENT sont supposés être utilisés à la fois pour les tâches pédagogiques et informatives envers les parents, nous nous sommes attachés à étudier précisément comment les enseignants considèrent l'ENT pour ces types d'usage. Quelle est la nature des tâches qui nécessitent la mobilisation de l'ENT ? Vont-ils l'utiliser pour transmettre des informations aux parents ou, au contraire, vont-ils privilégier uniquement une utilisation pédagogique en classe à destination des élèves ?

Dans la deuxième étude, nous faisons l'hypothèse que les enseignants du primaire vont rencontrer certains obstacles dans l'appropriation de l'outil. On se demande donc quels sont les facteurs défavorables à l'utilisation de ONE et quels sont les facteurs en favorisant l'utilisation.

Compte tenu de la nouveauté du projet ENT dans le primaire, nous nous situons pour les deux études dans une démarche inductive et exploratoire et non dans une démarche hypothétique. Nous allons donc partir des observations et des données issues du terrain afin de formuler des recommandations et des directions d'étude.

Pour décrire et analyser les différents facteurs liés à l'usage des ENT en contexte scolaire, nous proposons d'utiliser des modèles de l'acceptation et de l'appropriation technologique.

### 1.1. Une perception de l'acceptation de l'ENT

Parmi les différents modèles permettant d'évaluer l'acceptabilité technologique, le modèle du TAM (Technology Acceptancy Model) de Davis (Davis, 1989) est sans nul doute l'un des plus utilisé. Il cherche à identifier les intentions ou le maintien d'usage de dispositifs par la mesure de l'utilité perçue et la facilité d'utilisation perçue. Pour autant, comme l'indique Brangier *et al.,* (Brangier *et al.,* 2009), cette approche se révèle peu adaptée pour affiner la conception et analyser l'implémentation d'un système, car elle ne considère pas les pratiques effectives à l'œuvre. Plus précisément, différentes études (Bruillard, 2011, Bruillard et Hourbette, 2008) ont montré que le modèle TAM n'était pas adapté pour étudier l'acceptabilité des plateformes éducatives pour plusieurs raisons comme l'insuffisance méthodologique du modèle (structure factorielle qui n'est pas systématiquement répliquée, utilisation du questionnaire comme unique méthode d'évaluation) ou encore l'inadéquation au terrain éducatif. Le TAM apparaît donc comme un modèle prédictif et déterministe qui reste circonscrit aux facteurs sociocognitifs individuels et ne prend pas en compte le contexte d'utilisation de la technologie propre au milieu éducatif : cadre réglementaire, programme scolaire, relation avec les familles, histoire et pratiques professionnelles.

D'autres approches, d'inspiration ergonomique, ont traité la problématique de l'acceptabilité des Environnements Informatiques pour l'Apprentissage Humain (EIAH). Tricot *et al.,* (Tricot *et al.,* 2003) ont ainsi adapté le modèle de Nielsen au contexte scolaire. En effet, dans le modèle de Nielsen, l'acceptabilité et la décision d'utiliser une technologie dépendent de deux critères : sa facilité d'usage et son utilité pour les utilisateurs. Tricot montre que, dans le cadre scolaire, l'utilisation des technologies reste faible en dépit du fait qu'elles sont faciles d'utilisation et utiles pour

**Elena CODREANU, Christine MICHEL, Marc-Eric BOBILLIER-CHAUMON, Olivier VIGNEAU**

l'éducation. Tricot définit l'acceptabilité d'un EIAH comme « la valeur de la représentation mentale (attitudes, opinions... plus ou moins positives) à propos d'un EIAH, de son utilité et de son utilisabilité » (p. 396). Il propose dans son modèle d'évaluer conjointement l'acceptabilité, l'utilité et l'utilisabilité d'un l'outil à l'aide de tests utilisateurs ou d'inspections ergonomiques. Selon lui, ces trois dimensions sont complémentaires et liées par des relations. Ce modèle reste cependant circonscrit aux caractéristiques techniques de l'EIAH et aux facteurs d'utilisation, que Tricot appelle les aspects « pratiques ». Il ne prend pas en compte le contexte socioculturel du milieu d'implantation ou le rôle de la communauté (collègues ou hiérarchie) dans l'adoption d'une technologie, ce qu'il nomme les aspects « sociaux ». Aussi, pour appréhender ces différents aspects dans toute leur richesse et complexité, nous proposons d'utiliser la théorie de l'activité.

### 1.2. Une inscription de l'acceptation et de l'appropriation dans l'activité

Une troisième approche théorique est l'approche socio-constructiviste qui utilise le concept théorique d'appropriation. Pour comprendre l'appropriation d'une technologie, il faut se centrer sur son inscription dans un contexte social spécifique. La théorie de l'activité, détaillée par Engeström (Engeström *et al.,* 1999) propose de qualifier les éléments du contexte d'usage en considérant différents aspects intervenant dans la réalisation de l'activité. En effet, plutôt que de parler d'usage, la théorie de l'activité propose de parler de système d'activité : l'utilisateur a un objectif précis, le réalise en utilisant des instruments (outils) et s'inscrit socialement dans une communauté (l'ensemble de personnes qui interviennent dans une activité) elle-même liée à des règles de fonctionnement (les normes et les règles à respecter dans une activité), et une division du travail (la manière dont les rôles sont distribués entre les sujets). Les systèmes d'activité sont caractérisés par des contradictions (ou tensions internes) qui favorisent et déclenchent l'innovation ; ces changements sont source de développement. Ainsi la théorie de l'activité nous semble utile pour qualifier le contexte, mais aussi les dynamiques d'acceptation et d'appropriation technologiques.

Dans le cadre de notre recherche, le système d'activité des enseignants a pour objet les pratiques enseignantes quotidiennes. Elles sont réalisées avec ou sans instrument. En effet, la plupart des pratiques des enseignants ont un caractère pédagogique (envers les élèves) ou communicationnel (envers les parents), et sont complétées par l'utilisation d'instruments comme le tableau, des affiches, des cahiers (Karasavvidis, 2009). Ces pratiquent suivent les règles de fonctionnement propres au système scolaire et s'inscrivent dans une communauté éducative formée par les enseignants, les élèves, les parents. La division du travail décrit la pratique effective du métier et la répartition des tâches entre les acteurs. Dans l'éducation et le suivi des élèves, les enseignants et les parents travaillent ensemble, mais dans des contextes différents. Chaque acteur a ainsi sa part de responsabilités bien délimitée. Avec l'arrivée d'un nouvel outil technologique, qui sera utilisé à la fois en classe et à la maison, ces rôles et ces identités différentes peuvent entrer en conflit.

Selon Engeström (Engeström *et al.,* 1999), l'appropriation d'une technologie ne se fait pas sur un terrain neutre, mais dans un maîtrise signifie savoir utiliser différents moyens médiateurs sans difficulté, l'appropriation signifie prendre quelque chose qui appartient aux autres et le faire sien. Cela doit être interprété dans le sens d'une adaptation de l'outil à ses propres pratiques. L'appropriation est selon Jonsson (Jonsson, 2007), « le processus graduel à travers lequel les participants deviennent de plus en plus performants dans l'utilisation d'un outil » (p. 11). À la différence de la maîtrise, qui suppose l'acquisition d'une compétence, l'appropriation inclut, en plus d'une compétence technique, celle d'utiliser la technologie pour mener un travail innovant dans un contexte donné. L'appropriation serait donc profondément liée à la notion de changement. Utiliser un éditeur de texte à l'école ne change pas beaucoup les pratiques mais pouvoir faire des modifications sur un texte numérique sans avoir à le copier peut changer l'importance accordée traditionnellement à l'écriture. Ce sont l'usage et l'appropriation qui donnent de la valeur à la technologie. L'appropriation représente donc « la maîtrise de l'action du travail par celui qui l'exécute et les modifications induites dans ce qui avait été prévu par les concepteurs » (Bernoux, 2004, p. 56). Pour Rogoff (Rogoff, 1995), l'appropriation est « le processus selon lequel les individus transforment leur compréhension des activités à travers leur propre participation (à ces activités) » (p. 150). Ici, l'auteur parle surtout des outils culturels, comme le langage, les procédures et les technologies. L'approche instrumentale de Rabardel (Rabardel, 1995) va dans la même direction. Pour lui, l'appropriation représente le processus d'intégration d'un nouveau dispositif dans des pratiques et des manières de faire préexistantes. Pour Rabardel, un instrument ne se réduit pas au simple artefact technique, il contient en plus des modes opératoires qui sont

**Elena CODREANU, Christine MICHEL, Marc-Eric BOBILLIER-CHAUMON, Olivier VIGNEAU**

construits par l'humain, appelés schèmes d'utilisation. Les schèmes peuvent être préexistants ou nouvellement attribuées par l'utilisateur et peuvent mener à la modification de l'artefact. L'appropriation est le processus qui favorise la formation des schèmes d'utilisation et la transformation des artefacts en instruments au sein des activités humaines (Rabardel, 1995). Selon cet auteur, la genèse instrumentale représente la transformation d'un artéfact (constitué uniquement de la machine et du système d'opération) en instrument (constitué d'un artéfact et d'un schème d'utilisation) et représente une conceptualisation (au plan des instruments) des processus d'appropriation. Ainsi, les genèses instrumentales sont témoins du processus d'appropriation. Cette approche d'inspiration constructiviste suggère qu'il n'y a pas de relation unilatérale technologie-individu et que les deux s'influencent réciproquement : l'individu attribue du sens à la technologie en se l'appropriant et se laisse influencer à son tour par la technologie en modifiant ses pratiques.système complexe de pratiques préexistantes. Une technologie ne prend sens qu'au moment de l'interaction humaine qui la façonne et l'adapte continuellement. Wertsch (Wertsch, 1998) sépare la notion d'appropriation de celle de maîtrise d'un outil. Pendant que la

Nous constatons ainsi que la principale notion mobilisée est l'appropriation et il n'y a pas de distinction précise entre la notion d'appropriation et celle d'acceptation. Bobillier-Chaumon (Bobillier-Chaumon, 2016) considère que l'appropriation d'un outil technologique représente une condition préliminaire pour que l'outil soit accepté. Lorsqu'un individu s'approprie un outil, il y apporte une contribution qui lui permet de s'y reconnaître, de le « faire sien », de lui donner un sens et donc de l'accepter. L'acceptation située est ainsi définie « comme la façon dont un individu, mais aussi un collectif, une organisation, perçoivent au gré des situations quotidiennes les enjeux liés à ces technologies (atouts, bénéfices, risques, opportunités) et y réagissent (favorablement ou non) » (p. 362). Ainsi, l'adoption d'une technologie s'inscrirait sur un continuum allant des représentations a priori d'un système jusqu'à des formes d'acceptation construites sur la base des usages réalisés. Nous retrouverons une approche similaire chez Alter (Alter, 2000) qui signale que ce ne sont pas les technologies elles-mêmes qui ont de la valeur mais la capacité des utilisateurs à s'en servir de manière créative, en détournant l'usage. Selon cet auteur, c'est la création du sens qui compte le plus dans l'innovation. Les dirigeants devraient renoncer au caractère formel de la technologie et laisser aux utilisateurs des marges d'interprétation de celle-ci. Dans la même direction, Orlikowski (Orlikovski, 1992) affirme que la technologie seule dans une organisation n'a aucune signification. Une fois créée et introduite dans l'organisation, la technologie reste « non animée » jusqu'au moment où elle prend du sens dans l'usage et est manipulée directement ou indirectement.

Le tableau 1 présente une synthèse des différentes approches théoriques et notre propre choix de positionnement pour structurer l'analyse.

Sur cette base, nous proposons d'analyser conjointement l'appropriation et l'acceptation de l'ENT ONE par les enseignants en considérant respectivement les usages concrets et les représentations liées à l'ENT exprimés au cours d'entretiens.

**Tableau 1· Distinction entre les différentes approches théoriques**

| Approche théorique | Auteur | Explication de l'acceptation et de l'appropriation | |
|---|---|---|---|
| TAM | Davis, 1989 | L'acceptation est définie par les perceptions des utilisateurs. | |
| Ergonomique | Tricot, 2003 | L'acceptation dépend de l'utilisabilité et l'utilité du système. | |
| Socio-constructiviste | Rabardel, 1999 | L'appropriation est le processus qui favorise la formation des schèmes d'utilisation et la transformation des artéfacts en instruments au sein des activités humaines. | |
| | Rogoff, 1995 | L'appropriation est la transformation des activités. | |
| | Bernoux, 2004, | La maîtrise de l'outil de travail et les innovations apportées. | |
| | Bobillier-Chaumon, 2016 | Différenciation entre appropriation (vue comme la maîtrise d'un outil) et l'acceptation (vue comme la décision de continuer à utiliser l'outil). Les deux processus s'inscrivent dans une situation d'usage. L'appropriation est antérieure à l'acceptation (Bobillier-Chaumon, 2016). | ➔ **Nous avons retenu ce modèle et cette explication** |

## 2. Etude de l'appropriation de ONE

Cette étude a pour objectif d'identifier et analyser les types et volumes d'usage des utilisateurs de l'ENT, à savoir les enseignants, les enfants et les parents, après 6 mois à 1 an d'utilisation, de manière à mieux appréhender

**Elena CODREANU, Christine MICHEL, Marc-Eric BOBILLIER-CHAUMON, Olivier VIGNEAU**

leurs modalités d'appropriation et de mieux comprendre ce qui les sous-tend. L'appropriation sera évaluée à travers l'étude des usages des différents services de l'environnement numérique (nature, volume et fréquence) et aussi des usages innovants qui ont pu être développés avec l'ENT ONE.

### 2.1. Description de l'outil

L'ENT ONE utilisé dans cette étude a été conçu spécialement pour le premier degré en respectant les principes ergonomiques des interfaces adaptées aux enfants (Budiu et Nielsen, 2010), (Lueder et Rice, 2007). Ainsi, l'interface de ONE paraît simple, intuitive et attrayante (voir figure 1). Les fonctions de collaboration proposées sont la messagerie, le blog et l'espace documentaire. ONE propose de plus, des fonctions de personnalisation (mon compte, mon humeur), de notification (fil de nouveautés, anniversaire) et d'organisation (calendrier) et un site d'école. Chaque utilisateur a la possibilité de personnaliser son profil avec une photo et des informations personnelles (devise, humeur du jour, informations sur les préférences en termes de loisirs, cinéma, musique, alimentation). Les élèves sont inclus par défaut dans le groupe constitué de leur classe et ont accès aux contenus publiés dans ce groupe par leur enseignant.

L'ENT a été implémenté dans 12 écoles des Académies de Versailles et de Caen pendant une phase d'expérimentation de 2 années scolaires (2014-2015 et 2015-2016). Il s'agit d'une étape obligatoire avant la généralisation de l'ENT à un territoire plus vaste. Il n'y avait pas à ce moment-là d'injonction de la part de l'Etat concernant l'utilisation des ENT dans l'enseignement primaire. La participation à cette étape d'expérimentation s'est donc faite volontairement. La phase d'expérimentation est une étape menée en collaboration entre l'éditeur informatique (Open Digital Education), les représentants de l'Etat et du Ministère de l'Education c'est-à-dire les IEN (Inspecteurs de l'Éducation Nationale), les collectivités locales (mairies et communautés de communes) et les acteurs finaux (directeurs d'écoles, enseignants). Les conditions de déploiement de la phase d'expérimentation ont été négociées entre ces acteurs. Dans notre cas, dans chaque école un référent ENT a été nommé parmi les enseignants. Ce référent était formé à l'ENT et pouvait former ensuite ses collègues. L'accompagnement et la formation étaient assurés par les animateurs TICE[1] des écoles et/ou par l'équipe technique de Open Digital Education. Le type d'accompagnement nécessaire à la phase d'expérimentation (réalisée à une échelle plus restreinte) est plus important que celui nécessaire à la phase de généralisation car l'outil est en continuelle évolution ; son fonctionnement peut changer pendant cette phase où de nouveaux services peuvent être proposés.

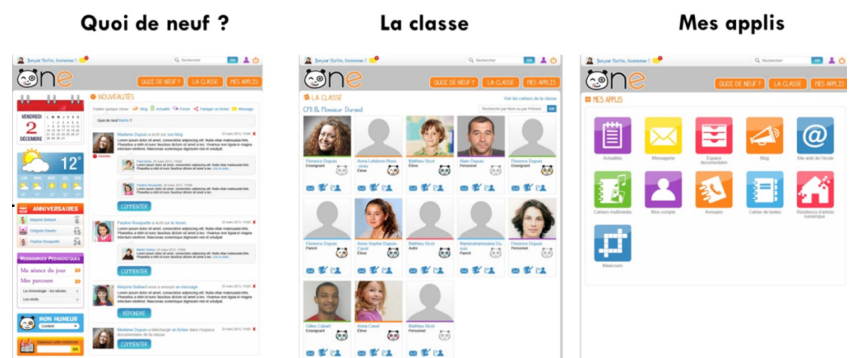

**Figure 1 · Interfaces des pages « Fil de nouveauté » « La classe » et « Mes appli » de l'ENT ONE**

### 2.2. Collecte des données

Au moment de l'étude, deux services de l'ENT ONE étaient déployés sur l'ensemble de la population d'utilisateurs observé : le cahier multimédia et le blog. L'étude d'usage est donc centrée sur ces deux fonctionnalités. Ces deux services se ressemblent par le fait que les utilisateurs peuvent y publier du contenu, comme des articles autour de la vie de la classe, de leçons, des textes rédigés par les élèves eux-mêmes. Ils sont tous les deux utilisables pour répondre à des besoins de type pédagogique ou communicationnel. Par exemple, avec les deux services il est possible de publier des contenus à visées pédagogiques (comme des leçons, des exercices) ou des contenus à visées informationnels à destination des parents (comme les photos des activités sportives, le journal de vie de la classe). Le blog et le cahier permettent de publier des contenus multimédias c'est-à-dire combinant du texte, des images, des vidéos ou des fichiers audio (voir figures 2 et 4). Les publications des élèves sont modérées et corrigées. En effet, la fonction « Validation de billets soumis » permet aux enseignants de lire et corriger les contributions des élèves avant de valider leur publication. La différence entre le blog et le cahier multimédia est de forme : le blog permet de publier des billets, alors que le cahier permet de publier des pages, qui sont visibles avec un défilement similaire au cahier papier.

**Elena CODREANU, Christine MICHEL, Marc-Eric BOBILLIER-CHAUMON, Olivier VIGNEAU**

Nous avons pu accéder aux blogs et cahiers multimédias des classes par l'intermédiaire d'un compte invité, ouvert avec l'autorisation des enseignants et de la direction, dans chaque école étudiée. En effet, les ENT fonctionnent sur la base de comptes personnels ayant une authentification unique. Aucune consultation n'est possible sans cet accès. Nous avons observé les publications réalisées sur l'ENT ONE par 26 enseignants et 71 élèves de 12 écoles différentes des Académies de Caen et Versailles entre septembre 2014 et juin 2015. Ce corpus d'analyse est composé de 1102 publications (voir tableau 2). Notre étude concerne donc, à ce stade, uniquement les activités de publication. Pour évaluer l'appropriation à partir des activités de consultation, nous avons prévu d'analyser dans une étude ultérieure les statistiques de connexion à la plateforme.

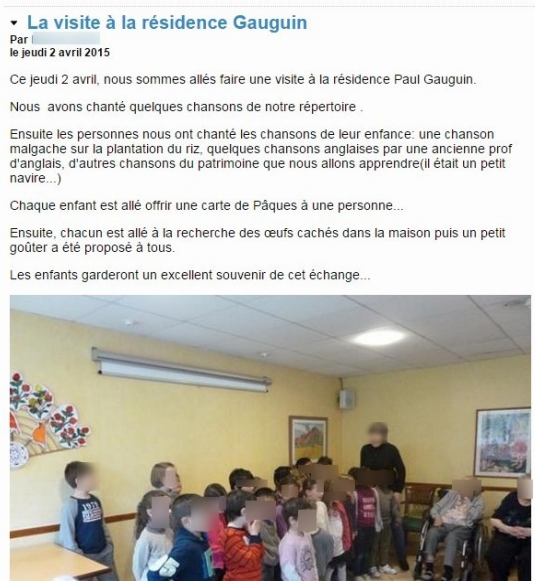

**Figure 2 · Exemple de thématique communicationnelle sur le Blog**

### 2.3. Critères d'analyse

Les publications ont été catégorisées en fonction de l'auteur, de son cycle d'apprentissage et du thème de la publication. Les auteurs sont de trois types : enseignant, élève, parent. Les cycles d'apprentissage sont de trois types également : C1, C2, C3. En effet, l'école primaire est organisée en trois cycles pédagogiques (Ministère de l'Education Nationale, 2013) : le cycle des apprentissages premiers (Cycle1) constitué de l'école maternelle ; le cycle des apprentissages fondamentaux (Cycle 2) qui commence avec le cours préparatoire (CP) et se poursuit avec le cours élémentaire première année (CE1) et deuxième année (CE2) et le cycle de consolidation (Cycle 3) qui s'étend du cours moyen première année (CM1) jusqu'à la classe de sixième qui est dans le secondaire.

Une première lecture de l'ensemble des publications faites sur les blogs et cahiers accessibles a permis d'identifier 5 thèmes de publications : communicationnelles internes, communicationnelles externes, pédagogiques individuelles, pédagogiques collaboratives et loisirs. Elles seront décrites plus précisément dans la suite.

### 2.4. Résultats

#### 2.4.1. Volumes et auteurs des publications par type de service

Le tableau 2 décrit la distribution de publications faites par les enseignants et les élèves sur le blog et le cahier multimédia. On constate que sur le nombre total de 1102 publications, 946 sont effectuées sur le service Blog et 156 sur le Cahier Multimédia. Les enseignants publient plus massivement sur le blog que sur le cahier (796 vs. 44) alors que les volumes sont plus équilibrés pour les élèves (150 vs. 112). Les parents n'ont fait aucune publication.

**Tableau 2 • Distributions de publications sur le Blog et le Cahier Multimédia**

| Type de service | Auteur | | | Total |
|---|---|---|---|---|
| | Elèves | Enseignants | Parents | |
| Blog | 150 | 796 | 0 | 946 |
| Cahier multimédia | 112 | 44 | 0 | 156 |
| Total | 262 | 840 | 0 | 1102 |

L'absence de publication des parents peut s'expliquer par le fait que dans la quasi-totalité des écoles étudiées, les enseignants avaient choisi de ne pas ouvrir aux parents le droit de publier du contenu sous forme d'article de blog ou de cahier multimédia. Les enseignants considèrent

**Elena CODREANU, Christine MICHEL, Marc-Eric BOBILLIER-
CHAUMON, Olivier VIGNEAU**

que le rôle des parents doit se limiter à la consultation des informations publiées sur l'ENT et ne doivent pas avoir la fonction de contributeur actif. Ce constat a été fait suite à notre immersion sur le terrain et la connaissance du milieu. Lorsque nous aurons analysé les statistiques de connexion, nous pourrons apporter des éléments complémentaires de discussion sur l'implication des parents en regard du rôle que les enseignants leur ont assigné.

### 2.4.2. L'analyse de thématiques de publications

L'analyse thématique des publications a donc révélé 5 types de publications : (1) communicationnelles externes, (2) communicationnelles internes, (3) pédagogiques individuelles, (4) pédagogiques collectives et (5) loisirs.

Le thème 1 correspond à des publications relatives aux activités réalisées à l'extérieur de l'école comme des photos et descriptifs sur les sorties cinéma, piscine, les voyages scolaires, les visites au collège, maison de retraite, etc. Le thème 2 correspond à des publications en lien avec les activités réalisées dans la classe ou à l'intérieur de l'école (comme la photo du cours de gymnastique, les photos des fêtes d'anniversaire). Le thème 3 correspond à des publications liées à des activités pédagogiques individuelles réalisées avec ou via l'ENT comme la rédaction de textes par les élèves, la publication de vidéos éducatives, la publication de comptines, de dessins d'exercices de mathématiques ou de grammaire (voir figure 4). Le thème 4 correspond à des publications équivalentes mais pour des activités faites en collaboration avec d'autres élèves (comme des projets de classe réalisés par groupe d'élèves). Le thème 5 correspond à des publications qui décrivent des activités extrascolaires liées à la musique, le sport, des voyages, etc. Les thèmes 1 et 2 sont à destination des parents. Les thèmes 3, 4 et 5 sont à destination des enseignants et/ou des élèves, mais peuvent également être ouverts aux parents.

La figure 3 décrit le nombre de thématiques couvertes par auteur, cycle et type de services (Blog ou Cahier Multimédia) et le tableau 3 récapitule les résultats obtenus, en volume et pourcentage par cycle et type d'auteur.

**Tableau 3 • Distribution des publications en fonction du cycle d'apprentissage, de la thématique et de l'auteur.**

| Auteurs (Cycles) | Thématiques de la publication | | | | | |
|---|---|---|---|---|---|---|
| | Com. ext. | Com. int. | Péd­ag. ind. | Péd­ag. col­lec. | Loi­sirs | To­tal |
| **Cycle 1** | 23 *(7 %)* | 308 *(91,5 %)* | 2 *(0,5 %)* | 4 *(1 %)* | 0 *(0 %)* | 337 |
| **Cycle 2** | 51 *(16 %)* | 252 *(77 %)* | 13 *(4 %)* | 10 *(3 %)* | 0 *(0 %)* | 326 |
| **Cycle 3** | 87 *(20 %)* | 102 *(23 %)* | 81 *(18 %)* | 104 *(24 %)* | 65 *(15 %)* | 439 |
| **Total en­seignants** | 114 *(14 %)* | 634 *(75 %)* | 76 *(9 %)* | 16 *(2 %)* | 0 *(0 %)* | 840 |
| **Total élèves** | 47 *(18 %)* | 28 *(10 %)* | 20 *(8 %)* | 102 *(39 %)* | 65 *(25 %)* | 262 |
| **Total** | 161 *(15 %)* | 662 *(60 %)* | 96 *(9 %)* | 118 *(11 %)* | 65 *(5 %)* | 1102 |

Le tableau 3 montre que 75 % des publications totales sont à vocation communicationnelle, alors que 20 % sont à destination pédagogique et seulement 5 % à destination de loisirs. De plus, les publications sont réali­sées principalement par des enseignants (840) et moins par des élèves (262) alors qu'unitairement les enseignants sont moins nombreux (34) que les élèves (71). L'appropriation de ONE se fait donc principalement par le biais de la communication à destination des parents et des élèves. Cela correspond également à ce que nous avons constaté pendant les discus­sions informelles réalisées avec les enseignants. Au début, lorsque les en­seignants observent pour la première fois l'ENT, les premiers usages qu'ils se représentent concernent la mise en ligne de photos des activités réali­sées à l'école à destination des parents. Sur ces usages communicationnels se greffent, plus tard, des usages pédagogiques : mise en ligne des comp­tines, des vidéos éducatives, des consignes d'exercices visant à impliquer les élèves et les inciter à rédiger leurs propres textes.

**Elena CODREANU, Christine MICHEL, Marc-Eric BOBILLIER-CHAUMON, Olivier VIGNEAU**

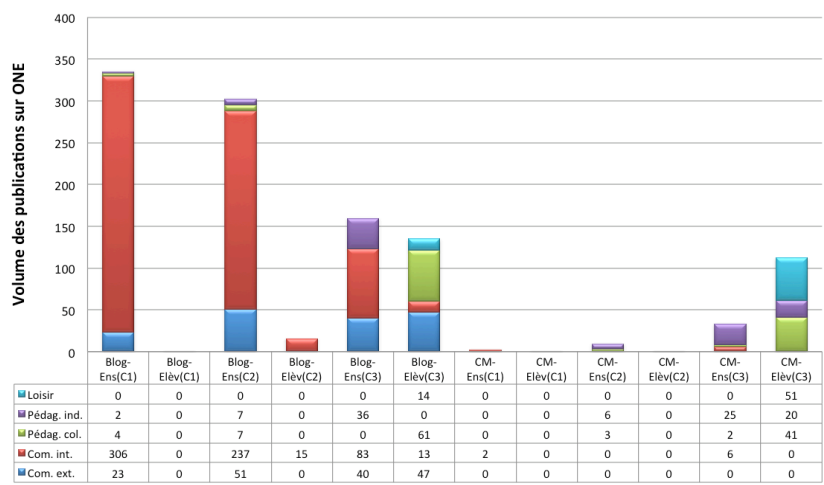

**Figure 3 · Volumes des thématiques par auteur, cycle et type de services (Blog et Cahier Multimédia)**

L'analyse de la distribution des thématiques par cycle à partir du tableau 3, montre que 91,5 % des publications faites dans le cycle 1 sont de type information école, 7 % des publications font référence à des activités extérieures à l'école, seulement 1 % concerne des publications pédagogiques. Presque toutes les publications du cycle 1 sont faites sur le blog (voir figure 3). Toutes les publications sont faites par les enseignants. Les élèves ne sont pas suffisamment autonomes et ne maîtrisent pas l'écriture et la lecture à l'école maternelle.

Dans le cycle 2 (CP, CE1 et CE2), les enseignants restent encore les principaux auteurs avec 95 % des publications réalisées. Quelques élèves commencent à publier, en particulier sur le blog. La proportion de publications communicationnelles relatives aux activités internes diminue légèrement (à 77 %), laissant la place à des informations relatives aux sorties scolaires (15 %). La proportion de publications de nature pédagogique reste également assez limitée (seulement 7 % en total). Ces publications correspondent pour les enseignants à des leçons, des comptines, des vidéos éducatives et pour les élèves à des textes rédigés à la demande des enseignants. La connaissance du terrain nous montre que les activités réalisées avec l'ENT sont souvent optionnelles. L'intérêt pour les enseignants est de familiariser les élèves à l'ENT et leur donner des moyens de l'expérimenter simplement, plutôt que l'utiliser pour réaliser des projets pédagogiques complexes. À titre d'exemple, une enseignante de CP a réali-sé un cahier multimédia musical pour recueillir des comptines d'autrefois fournies par des parents d'élèves et des grands parents et a ensuite organisé des rencontres musicales avec les parents et grands-parents volontaires pour chanter avec les élèves. Il s'agissait d'un projet intergénérationnel et collaboratif, où les élèves pouvaient apprendre et chanter ces comptines disparues, accompagnés par des aînés. L'ENT a été utilisé pour recueillir des notes musicales, le texte des comptines mais aussi des fichiers audio. Ces supports ont été rendus accessibles à tous les acteurs, enseignants, élèves, parents, en classe comme à la maison.

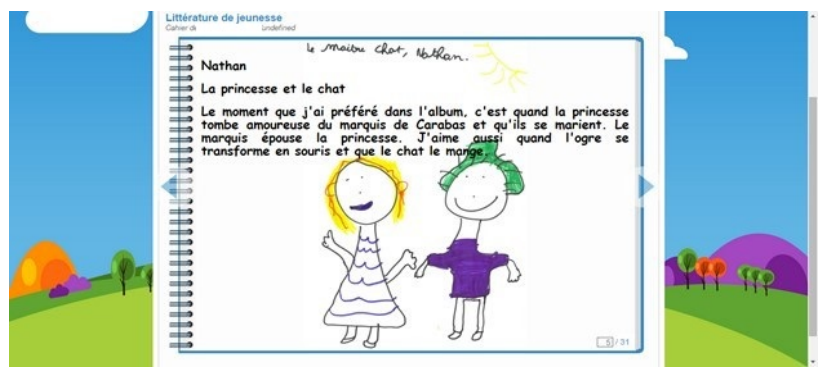

**Figure 4 • Exemple de thématique pédagogique (rédigé par un élève)**

Dans le cycle 3 (CM1 et CM2), les proportions changent. La place des élèves augmente, car ils produisent 56 % des publications totales du cycle 3. Les pourcentages de publications communicationnelles baissent à 23 % (pour les informations internes) et 20 % (pour les externes). À l'inverse, la proportion de publications pédagogiques augmente à 18 % pour les activités individuelles et à 24 % pour les activités collaboratives. De plus, 15 % de publications sont relatives aux loisirs alors que ce type était complètement absent des cycles 1 et 2. Rappelons que ces publications de type Loisirs sont faites à l'initiative des élèves et non pas à la demande des enseignants, dans le but de partager avec les autres élèves des informations sur leurs activités extrascolaires sportives, familiales ou culturelles (conformément à notre connaissance du terrain). Les élèves étant plus autonomes à ce niveau, ils arrivent à maîtriser la plupart des fonctionnalités de l'ENT et peuvent donc respecter les consignes des enseignants pour rédiger en relative autonomie des publications, soit sur le cahier multimédia, soit sur le Blog.

**Elena CODREANU, Christine MICHEL, Marc-Eric BOBILLIER-CHAUMON, Olivier VIGNEAU**

Les élèves de cycle 3 réalisent une partie des tâches précédemment réservées aux enseignants comme la publication du journal de vie de la classe, avec des informations sur les activités en classe ou extérieures (sorties cinéma, voyages scolaires). D'ailleurs, l'observation en contexte nous a permis de comprendre que les enseignants attribuent une connotation pédagogique à quasi toutes les publications réalisées par les élèves. Par exemple, lorsqu'un élève rédige un billet de blog à destination des familles qui décrit une sortie scolaire, cette activité permet aussi à l'élève de développer des aptitudes spécifiques comme la capacité d'utilisation d'un outil numérique, la capacité de structurer correctement un texte, de travailler l'orthographe, le résumé, etc. Ce type de rédaction a donc un rôle initial communicationnel et reçoit ultérieurement un deuxième rôle, pédagogique. Ce type d'activité à double connotation, scolaire et communicationnelle, représente un usage original. C'est un exemple typique de nouvelle pratique réalisée à l'aide de la technologie.

Une caractéristique est commune à toutes les publications, toutes les classes et tous les niveaux confondus : la dimension ludique et humoristique. En effet, les élèves, comme les enseignants, créent des blogs ou des cahiers autour des sujets « moins sérieux » comme des blagues, des charades, des informations insolites, des photos d'animaux, des concours de mathématiques interclasses comportant une « énigme » à résoudre. Ce type de publications, autour du jeu, du rire, de la compétition ludique sont des modes très favorables d'appropriation de l'ENT car ils favorisent le partage et l'intégration de l'outil à la culture et aux pratiques scolaires de l'école primaire. L'appropriation de l'outil se fait donc aussi par des phases d'expérimentation où les enseignants vont avec les élèves inventer ou adapter de nouvelles possibilités offertes par l'ENT et juger comment il peut répondre aux besoins spécifiques des enfants les plus jeunes. Les blogs et les cahiers créés autour de ces sujets deviennent un miroir « numérique » de la classe, une représentation des expériences et des émotions vécues en classe à travers le choix des textes et des images publiées et commentées. Ces observations nous permettent de conclure que les enseignants et les élèves ont adapté l'ENT et ses fonctionnalités à leurs pratiques scolaires.

Le fait qu'un quart des publications faites au cycle 3 soient collaboratives (104 sur 439, voir tableau 2) signifie que les enseignants accordent une place importante aux tâches faites en groupe (comme le cahier de voyage réalisé à deux). Les enseignants préfèrent faire travailler les élèves ensemble pour leur permettre de s'entraider et aussi dans une logique de transmission des connaissances informatiques. Ainsi, les élèves moins indépendants peuvent apprendre à utiliser l'ENT avec l'aide d'un autre élève « expert ». Ces nouvelles formes de collaboration sont réalisées en classe mais aussi à distance. En effet, le cahier de voyage est complété par les élèves à la maison ou à l'école. Ce type d'usage est également innovant car la collaboration à distance n'était pas possible auparavant.

Nous avons donc vu trois exemples d'usages innovants. L'ENT représente pour les enseignants une nouvelle façon d'enseigner, d'informer et de renforcer les compétences des élèves.

### 2.4.3. Bilan sur l'étude de l'appropriation de ONE

Nous retenons de cette étude une appropriation principale de l'ENT ONE à travers la possibilité de communication avec les parents. Ainsi, les enseignants peuvent mettre beaucoup plus de photos que dans les cahiers papiers. Ils peuvent aussi poster des fichiers multimédias comme les vidéos des élèves. Ces résultats sont intéressants car ils semblent contredire les études qui affirment une non-intégration de la part des enseignants des pratiques communicationnels à destination des familles (Bruillard, 2011). Les enseignants refuseraient ces pratiques communicationnelles tournées vers les parents car ils craignent une implication trop forte de certains parents dans la vie de l'école, une intrusion dans leurs propres pratiques et une critique des personnes non-enseignantes et donc illégitimes à leurs yeux (Monceau, 2009). L'ENT ONE aurait donc un rôle positif, car il favorise cette tâche normalement refusée. On peut donc conclure que les enseignants du primaire semblent favorables à l'ouverture de l'école auprès des familles.

Dans un second temps, l'appropriation est construite à partir des usages pédagogiques comme la publication des comptines, des exercices ou des textes rédigés par les enfants eux-mêmes. Ce type d'usage est intéressant car il correspond : à la publication de contenus à double connotation (communicationnelle et pédagogique) et à la possibilité de faire publier alternativement les enseignants et les élèves. La distinction entre le travail de l'enseignant et celui de l'élève est, pour ce cas précis, effacée. De plus, une publication faite sur l'ENT est une forme de valorisation collective car elle est visible par l'ensemble des parents, d'élèves et les autres enseignants.

**Elena CODREANU, Christine MICHEL, Marc-Eric BOBILLIER-CHAUMON, Olivier VIGNEAU**

Si l'on considère que l'appropriation se fait lorsqu'une technologie devient « la sienne », lorsqu'elle s'intègre dans le contexte spécifique de travail des utilisateurs (Bobillier-Chaumon, 2016), (Engeström *et al.,* 1999) on peut constater que les différentes dimensions du contexte éducatif ont été touchées. En effet, les utilisateurs se sont approprié l'outil pour répondre à des objectifs de pédagogie, de transmission de connaissances, et de communication avec les parents. L'ENT est devenu un reflet fidèle et un support pour les activités pédagogiques, récréatives et ludiques réalisées en classe. L'appropriation de l'ENT s'est faite par des phases d'expérimentation successives. Les enseignants et les élèves ont expérimenté plusieurs types d'activités sur l'ENT. Certaines se sont pérennisées (comme les cahiers collaboratifs) alors que d'autres ont été abandonnées (comme les énigmes mathématiques) en fonction des expériences vécues et de la valeur attribuées par les initiateurs. L'appropriation de l'ENT ONE a été effective car les utilisateurs ont eu la possibilité d'imaginer des activités et de les tester en contexte et ainsi d'avoir un jugement sur celles qui sont les plus efficaces et valorisantes pour eux.

Cette étude a montré les pratiques des enseignants avec l'ENT. Dans l'étude suivante nous avons évalué, à l'aide d'entretiens, le ressenti des enseignants et les éventuelles tensions liées à l'utilisation de ONE.

## 3. Etude de l'acceptation de ONE

L'étude de l'acceptation a été centrée sur les déclarations des enseignants autour de l'utilisation de l'ENT. Elle a permis d'identifier le ressenti ainsi que le vécu des enseignants vis-à-vis de la technologie et les éventuels problèmes rencontrés. L'objectif était d'affiner l'interprétation des usages et de faire des recommandations d'amélioration à la fois sur la conception de l'outil mais aussi de son déploiement, en particulier concernant la formation et l'accompagnement. L'acceptation est évaluée à travers l'identification dans le discours des enseignants des situations et/ou des pratiques médiatisées qui peuvent montrer soit un effet favorable et bénéfique de ONE sur l'activité (en termes de maintien, de développement, de valorisation, d'innovation), soit au contraire un effet négatif et défavorable sur leur travail (empêchement, restriction, contraintes, dépréciation).

Notre démarche est essentiellement qualitative. Elle repose sur des entretiens semi-directifs (intégralement retranscrits) réalisés auprès des enseignants de manière à appréhender leur activité professionnelle ainsi que leurs différentes contraintes dans l'exercice de leur métier et dans l'usage de l'ENT pour leurs diverses pratiques. Nous avons cherché aussi à recueillir leurs représentations et attentes vis-à-vis du dispositif. Ces corpus ont fait l'objet d'un traitement systématisé selon la méthode d'analyse thématique de contenu (Bardin, 1996).

### 3.1. Milieu d'implantation

Tous les participants de notre étude font partie des Académies de Versailles et Caen (6 écoles de l'Académie de Versailles et 6 écoles de l'Académie de Caen). Ils ont accepté de manière volontaire d'expérimenter l'ENT ONE pendant 2 ans. Au moment de notre étude 26 enseignants (sur les 2 académies) s'étaient déclarés volontaires pour faire l'expérimentation et avaient utilisé l'ENT ONE sur une période allant de 3 à 6 mois. Ces enseignants sont un échantillon de ceux dont les publications ont été analysées dans l'étude précédente. Les deux études ont été réalisées sur la même période de temps, l'année scolaire 2014-2015.

### 3.2. Participants

Huit enseignants ont été interrogés durant 4 entretiens individuels et lors de 2 entretiens collectifs (avec 2 enseignants chacun). Parmi les enseignants, 2 sont directeurs d'école et assurent des cours (en CP et CM2). Les autres enseignants interviennent dans les classes de CP (2), CE1 (1) et CM2 (3). Le groupe de participants est composé de sept femmes et d'un homme. Les écoles sont toutes situées en milieu urbain, dans l'Académie de Versailles (6) et de Caen (2). L'âge moyen des répondants était de 46 ans, ils ont une ancienneté moyenne dans la structure de 8 années et une ancienneté dans la profession de 16 années. La relation avec les parents se fait traditionnellement par le cahier de correspondance papier, au téléphone ou en face à face. Tous sont par ailleurs des utilisateurs novices dans l'ENT : ils n'ont jamais utilisé auparavant de dispositif équivalent dans le cadre de leur enseignement. Ils ont cependant tous utilisé les ordinateurs pour préparer les leçons et certains ont un usage fréquent du tableau numérique interactif en classe (5 sur 8 enseignants).

### 3.3. Collecte des données

Chaque enseignant a participé à un entretien semi-directif. Les entretiens, d'une durée d'1h30 en moyenne, abordaient les thèmes suivants : expérience avec les TICE (Technologies d'information et de Communication pour l'Éducation), équipement informatique de l'école, représenta-tion de l'ENT, besoins par rapport à l'ENT, utilité de l'ENT, facilité d'usage et intention d'usage, difficultés d'utilisation, conditions d'exercice et d'implication dans la profession d'enseignant. Ayant la possibilité de parler librement, ils ont pu donner un avis critique sur les usages réalisés, leurs représentations de l'outil ou des fonctionnalités en développement comme le cahier de textes, le cahier de liaison numérique et le cahier multimédia. Ils ont également été invités à relater, avec la méthode des incidents critiques de Flanagan (Flanagan, 1954), des épisodes plutôt difficiles ou plutôt faciles dans l'usage de l'ENT.

### 3.4. Analyse des entretiens enseignants

Les entretiens ont été enregistrés et retranscrits en intégralité pour être analysés thématiquement. L'unité d'analyse des retranscriptions est la proposition (considérée comme une unité syntaxique élémentaire construite autour d'un verbe). Toutes les propositions ont été identifiées dans chaque phrase comme dans l'exemple suivant : « *Moi je leur ai montré comment faire des dossiers (proposition 1)/, mais c'est difficile pour les élèves (proposition 2)* ». Nous avons également distingué les commentaires plutôt favorables de ceux exprimant plutôt des tensions ou contradictions. Dans l'exemple précédent, la première proposition a été comptée comme remarque favorable (initiative d'accompagnement) et la deuxième comme remarque défavorable (difficulté d'usage). Nous avons fait des dénombrements et des calculs de pourcentage pour hiérarchiser les facteurs. Nous avons considéré que les utilisateurs ont accepté l'ENT lorsqu'ils évoquent les usages réussis qu'ils en font, les ajustements mis en œuvre ou les contradictions rencontrées et surmontées. Aucun thème n'était préétabli, nous n'avons retenu que les thèmes évoqués au moins trois fois.

### 3.5. Résultats

L'analyse a mis en évidence 4 thèmes principaux (voir tableau 4) et 16 sous-facteurs (voir tableau 5) : (1) facteurs liés à la pratique du métier (charge de travail, responsabilisation aux usages numériques, valorisation du travail), (2) facteurs relatifs au suivi pédagogique (pédagogie, sécurité et santé, émotions et attractivité) ; (3) facteurs relatifs à l'organisation sociale et du travail (collaboration, communication, réorganisation des pratiques de communication), (4) facteurs relatifs à l'usage et au déploiement de l'outil (facilité d'usage, utilité, feedback, équipement informatique et réseau, accompagnement). Nous présentons dans un premier temps les résultats des facteurs principaux, puis ceux des sous-facteurs.

### 3.5.1. Facteurs principaux.

Le tableau 4 indique que les facteurs liés à l'organisation sociale représentent le plus d'évocations positives (88), ce qui signifie que l'usage de l'ENT a un rôle important dans la communication et la collaboration à l'intérieur du système scolaire. À l'inverse, les facteurs liés au métier d'enseignant et ceux liés à l'usage et au déploiement de l'outil rassemblent le plus d'appréciations négatives. Le déploiement et l'usage de l'ENT semblent donc soulever d'une part, des questions sur la reconnaissance professionnelle et la pratique du métier d'enseignant et, d'autre part, des problèmes en matière d'adéquation aux usages scolaires. Nous présentons dans le paragraphe suivant une analyse par sous-facteurs (voir tableau 5) qui permet de préciser ces dimensions.

**Tableau 4· Occurrence des facteurs principaux.**

| Facteur | Nb appréciations positives | Nb appréciations négatives |
|---|---|---|
| **Métier** | 35 (15,56 %) | 90 (36 %) |
| **Suivi pédagogique** | 54 (24 %) | 57 (22,8 %) |
| **Organisation sociale** | 88 (39,11 %) | 14 (5,6 %) |
| **Usage et déploiement de l'outil** | 48 (21,33 %) | 89 (35,6 %) |
| **Total** | 225 (100 %) | 250 (100 %) |

### 3.5.2. Facteurs relatifs à l'exercice et la pratique du métier

Comme on peut le voir dans le tableau 5, la charge de travail perçue (induite par l'usage de l'ENT) apporte le plus d'évocations négatives (72). L'utilisation de ONE demande beaucoup de temps, surtout au début. Ils ont l'impression de devoir investir plus de temps pour maîtriser les fonctionnalités de l'ENT et pour trouver des applications intéressantes pour la classe. Ils ont aussi le sentiment que l'usage d'un ENT implique un travail soutenu et régulier, sur de nouvelles tâches qui ne relèvent pas directement de leur domaine de compétences : comme la prise de photos, les chargements sur l'ordinateur et ensuite sur l'ENT, la publication de billets de blogs, la rédaction de messages, la conception de projets pédagogique incluant l'ENT. Or, puisqu'ils ne disposent pas d'un temps scolaire spécialement dédié aux usages de ces technologies, ils sont alors obligés d'utiliser le temps de la pédagogie pour s'approprier ces outils. Le sentiment de charge de travail s'exprime aussi par une impression d'accroissement des sollicitations informationnelles. L'ENT s'ajoute, en effet, aux plateformes éducationnelles préexistantes : adresse mail académique, plateforme de gestion de carrière I-prof, plateforme de formation en ligne, plateformes didactiques, livret de compétences en ligne. Les enseignants se sentent ainsi continuellement submergés par une grande quantité de données à gérer (adresses mail, identifiants et mots de passe différents pour chaque plateforme, logique et fonctions des différents dispositifs,…) mais aussi par des contenus informationnels à traiter et à hiérarchiser (informations académiques, pédagogiques, événements à trier et à diffuser…). Face à la crainte de devoir faire un double travail avec l'ENT, certains refusent par exemple de publier les leçons sur l'ENT, car ils le font déjà avec leurs outils bureautiques « *je fais déjà la leçon sur le paper board, la remettre encore (sur l'ENT)… moi je n'ai pas envie de faire ça…* ».

Elena CODREANU, Christine MICHEL, Marc-Eric BOBILLIER-
CHAUMON, Olivier VIGNEAU

Tableau 5 · Occurrence des sous-facteurs.

| Sous-facteur | Nb appréciations positives | Nb apprécia-tions négatives |
|---|---|---|
| **Facteurs relatifs à l'exercice et la pratique du métier.** | | |
| Charge de travail | 0 | 72 |
| Responsabilisation aux usages numériques | 20 | 12 |
| Valorisation du travail | 15 | 15 |
| *Total* | *35* | *90* |
| **Facteurs relatifs au suivi des élèves.** | | |
| Pédagogie | 20 | 0 |
| Sécurité et santé | 4 | 57 |
| Emotions et attractivité | 30 | 0 |
| *Total* | *54* | *57* |
| **Facteurs relatifs à l'organisation sociale et du travail.** | | |
| Collaboration | 12 | 0 |
| Communication | 72 | 8 |
| Réorganisation des pratiques de communication | 4 | 6 |
| *Total* | *88* | *14* |
| **Facteurs relatifs à l'usage et au déploiement de l'outil.** | | |
| Facilité d'usage | 27 | 24 |
| Utilité | 9 | 6 |
| Feedback utilisateur | 4 | 39 |
| Equipement informatique et réseau | 0 | 6 |
| Accompagnement | 8 | 14 |
| *Total* | *48* | *89* |

La responsabilisation des élèves dans leurs usages numériques a été évoquée positivement 20 fois. Les enseignants pensent qu'ils ont un rôle à jouer dans la formation aux « usages responsables des outils numériques par les élèves ». D'autres considèrent en revanche que cette sensibilisation serait davantage du ressort des parents (12 évocations), d'une part parce que cela nuit à leur activité de formation, d'autre part parce que l'outil est massivement consulté à la maison par les enfants pour vérifier notamment leurs nouveaux messages. Pour ces raisons, le contrôle devrait relever davantage de la sphère privée. Ce que ne semblent pas partager les parents qui pensent au contraire que ce suivi doit être assuré par l'établissement qui le met à disposition. On voit ici que l'articulation ''école-maison'' nécessite une redéfinition des responsabilités et des attri-butions de chaque partenaire (parents et enseignants de parcours de formation) dans une division du travail mieux coordonnée (de contrôle et de suivi de l'usage).

La valorisation du travail ressort de manière positive au travers de 15 évocations. En fait, certains enseignants considèrent que l'ENT permet de mettre en évidence, via le blog, un travail de classe qui jusque-là était plutôt invisible ; comme les activités sportives, les sorties scolaires, les créations des élèves. Il devient ainsi un outil de reconnaissance du travail de l'enseignant ainsi que des productions des élèves. Mais cette reconnaissance du travail reste limitée par le fait que les parents sont peu impliqués dans le projet ENT et ne consultent que très rarement ces travaux (évocations négatives). Ce manque d'intérêt de la part des parents pour les informations mises à disposition provoque chez les enseignants une certaine déception voire un désir de ne plus utiliser l'ENT. C'est le cas de cette enseignante de CP qui a informé les parents via le cahier de correspondance de la tenue d'un cahier de vie de la classe sur l'ENT et les a invités à la consultation de ce cahier : « *J'ai un retour d'un parent qui m'a dit "Félicitations, super initiative" mais sinon il y en a qui n'ont pas signé le mot. C'est pour vous dire qu'ils ne sont pas intéressés. C'est un peu désespérant. Je n'ai vu que 3 élèves de la classe qui ont changé leur photo. Donc 3 sur 26 ce n'est quand même pas beaucoup. Si c'est pour qu'il n'y ait personne qui le regarde (le cahier), moi j'arrête de le remplir* ».

### 3.5.3 Facteurs relatifs au suivi des élèves

Selon les enseignants, l'intérêt principal des ENT vis-à-vis des élèves est d'aider à construire une relation plus attractive et stimulante, qui joue sur les émotions (30 évocations positives dans le tableau 4). L'ENT est motivant pour les élèves et fait apprécier le travail en classe. En matière de pédagogie, l'ENT est considéré comme un apport (20 évocations positives) pour la construction de l'expression verbale et de la communication des élèves ainsi que pour la responsabilisation et l'autonomie dans le travail avec ordinateur. Le blog est renseigné par les enseignants ou les élèves eux-mêmes. L'ENT est utilisé pour la réalisation des tâches pédagogiques, comme la publication des leçons, des vidéos éducatives, des textes rédigés par les élèves. Ces publications pédagogiques concernent toutes les matières : français, anglais, mathématiques, arts, sciences.

Les enseignants expriment en revanche de nombreuses craintes concernant la sécurité et la santé des enfants (57 évocations négatives contre

4 positives). En particulier, cela concerne les dérives (harcèlement, insultes) ou les détournements des outils de communication et de coordination. Les enseignants ont en effet un accès limité aux comptes des enfants et ils sont donc dans l'impossibilité de contrôler le contenu des messages échangés. Plusieurs enseignants ont alors créé un compte-élève fictif pour suivre et contrôler les échanges. Cela leur permet aussi de vérifier la qualité d'affichage des informations et des documents qu'ils publient sur l'ENT. On remarque que les enseignants, qui n'ont pas été en mesure de créer ce type d'usage innovant, sont moins satisfaits du dispositif. Ce type de détournement met en évidence l'intérêt de proposer aux enseignants des fonctions de surveillance des espaces et de vérification des publications. Une autre crainte concerne les transgressions d'usage de l'ENT par les enfants. Les enseignants déclarent notamment avoir des difficultés à authentifier les sources d'informations en provenance du système, comme cet exemple le montre *« moi j'ai reçu un message d'un parent, je ne sais pas si c'est le parent ou si c'est le grand frère qui a envoyé le message.../... du coup il faut que je repasse par le cahier de textes pour écrire un mot.../... sur le cahier de liaison il y a l'écriture, la signature, on connaît tout de suite rapidement la différence »*.

### 3.5.4 Facteurs relatifs à l'organisation sociale et du travail

L'ENT est particulièrement apprécié pour soutenir la communication avec les parents (72 appréciations positives). Certains enseignants créent des blogs et y font référence dans le cahier de liaison lorsqu'il y a de nouvelles informations à consulter. Les enseignants apprécient aussi le rôle positif que l'ENT joue sur la collaboration avec les autres enseignants (12 évocations). Le partage des ressources créées facilite l'organisation des activités ou des sorties en commun, ainsi que le travail pédagogique.

Les appréciations négatives (8) concernent surtout la communication via la messagerie qui ne distingue plus les temps scolaires et extra-scolaire. Ils évoquent l'intérêt de pouvoir paramétrer les horaires de transmission des messages des parents (pas de messages pendant le temps scolaire) ou entre les élèves. Vis-à-vis des parents, cela permettrait de limiter les messages intempestifs et de dernières minutes qui requièrent un travail supplémentaire durant le temps scolaire. Ils ont plus de contrôle par le cahier de liaison. Ces mesures peuvent être utiles au départ pour rassurer les enseignants et pour leur laisser le temps de mettre en place des actions de responsabilisation des élèves et des parents aux usages numériques.

### 3.5.5. Facteurs relatifs à l'usage et au déploiement de l'outil

Les enseignants ressentent une facilité d'usage (27 évocations positives) liée à la cohérence des fonctions et informations accessibles par le menu et les icônes. Les évocations négatives (24) portent sur des fonctionnalités de l'Espace Documentaire de l'ENT : ils souhaitent un partage de dossiers plutôt que le partage de fichiers « *les enfants reçoivent... [les fichiers] comme ça. Pour eux c'est pas facile, on a Documents Partagés et tout est mélangé ; musique, histoire. Si le nom du fichier il est un peu approximatif... ils savent pas* » et regrettent le manque de visibilité sur qui a consulté les contenus publiés et qui s'est connecté. En suivant le fil de nouveautés, les enseignants arrivent à connaître l'activité des autres utilisateurs (parents, élèves) si ces derniers modifient par exemple leur avatar ou la devise qui l'accompagne. Mais cet usage détourné de la consultation du fil ne marche pas pour les consultations simples, car elles ne laissent pas de trace. « *Ça c'est vrai que... s'ils ne changent pas leur humeur ou la devise, on ne le sait pas s'ils se sont connectés ou pas. Ça serait intéressant pour nous les utilisateurs de savoir qui a vu le contenu* ». Pour obtenir ces données, les enseignants font alors un travail supplémentaire qui consiste à envoyer un questionnaire via le cahier de correspondance, ou alors, ils demandent aux élèves si leurs parents se connectent. Cette traçabilité est importante pour construire une forme d'échange entre les différents partenaires du parcours de formation, afin de s'assurer que les informations publiées sont bien consultées et parviennent au destinataire. Sinon, il est difficile pour les enseignants de juger de l'utilité et de l'efficacité du système pour leur activité.

Le manque d'infrastructure informatique (matériels, réseau...) est par ailleurs considéré aussi comme un frein à l'acceptation de l'ENT (6 évocations). Les enseignants aimeraient utiliser l'ENT en classe avec les élèves mais ils manquent d'ordinateurs portables ou de tablettes. « *Il faudrait pratiquement avoir des ordinateurs en permanence dans les classes pour pouvoir vraiment l'utiliser dans la pédagogie de tous les jours* ». En effet, l'équipement technique des écoles est assuré par les communes qui choisissent d'investir ou non, ce qui peut créer des grandes inégalités entre les territoires. Comme les ENT sont des projets coordonnés par l'état et le Ministère de l'Éducation, cela peut créer des discordances sur le terrain qui peuvent éventuellement freiner les usages en classe. L'accès à l'ENT n'est pas équivalent non plus dans toutes les familles. Certaines offrent un accès continu à certains élèves et d'autres un accès restreint défini par les parents ou l'absence de connexion Internet. Enfin, les enseignants évoquent des manques d'accompagnement. Ils se considèrent mal formés aux usages de l'ENT. S'agissant d'une phase expérimentale d'implémentation, tous les moyens pour accompagner les enseignants n'ont pas été employés. Rappelons que dans chaque école un référent ENT a été désigné et a pu bénéficier de la formation. Il formait ensuite ses collègues. Ce fonctionnement a pu causer des difficultés en termes de transmission des compétences ou de réponses appropriées à toutes les problématiques soulevées par les collègues. À long terme, les responsables académiques devraient s'impliquer dans la formation et l'accompagnement de tous les enseignants. Sur ces aspects, des études supplémentaires seraient donc nécessaires pour comprendre comment s'opère la phase de généralisation des ENT dans les écoles primaires.

### 3.6. Bilan sur l'acceptation

L'étude de l'acceptation montre que les principales tensions sont liées à l'exercice et la pratique du métier, notamment la surcharge de travail, l'insuffisante connaissance des responsabilités et le problème de sécurité des élèves sur un espace de travail en ligne. Les facteurs relatifs à l'usage de l'outil posent aussi problème, par manque d'infrastructure, à la fois dans les écoles et dans l'espace domestique. Il nous semble nécessaire de renforcer l'accompagnement fait par les académies ainsi que la communication autour du numérique menée par le Ministère de l'Éducation. Le Schéma directeur des Espaces Numériques (Ministère de l'Education Nationale, 2012), document officiel des ENT, définit les rôles et les responsabilités de chaque futur utilisateur des ENT. Malheureusement, ce document est adressé surtout aux concepteurs et demeure peu lisible par le grand public qui ne possède pas ou peu de connaissances techniques sur ces plateformes numériques. Ces documentations devraient être vulgarisées pour être davantage accessibles aux enseignants, élèves et parents.

Les facteurs positifs de l'acceptation sont principalement ceux relatifs à l'organisation sociale, car l'ENT favorise la collaboration au sein de l'école (à travers le partage des contenus) et la communication avec les familles. De plus, l'ENT permet la valorisation du travail effectué en classe avec les élèves (textes rédigés, dessins, activités créatives) auprès des familles. Ce facteur positif est en quelque sorte atténué par la faible implication des parents. Cela veut dire que les enseignants peuvent trouver utile de passer par l'ENT pour informer les parents de la vie de l'école, cela ne signifie pas pour autant que les parents consultent vraiment les publications des enseignants. La faible implication des parents, déplorée par les enseignants de notre étude, est cohérente avec les problèmes du système d'enseignement actuel. En effet, même avant l'arrivée des ENT (qui sont supposés faciliter l'ouverture de l'école vers les familles), la communication entre l'école et les familles reste difficile et insuffisante (Bruillard, 2011). Les parents peuvent opposer une véritable résistance à la collaboration avec les enseignants, lorsqu'ils se voient nier leurs compétences parentales et reçoivent des messages « moralisateurs » à propos de leurs pratiques familiales, comme l'heure du coucher et l'alimentation (Monceau, 2009). Les parents se méfient donc du contact avec l'institution scolaire et évitent ainsi d'être enrôlés dans des activités collectives ou défendent tout simplement la séparation entre l'école et les familles justifiée par l'idée de « chacun sa place ». Ces problèmes sont plus fréquents dans les milieux sociaux défavorisés et rendent par conséquent la communication plus difficile. L'utilisation d'une technologie de type ENT ne peut pas résoudre ces problèmes préexistants dans le système d'enseignement. Nous recommandons d'encourager le dialogue entre parents et école, de manière à pouvoir poursuivre ces échanges par la suite avec des technologies telles que les ENT.

### 4. Discussion et conclusion

Les objectifs de l'article étaient de comprendre comment les enseignants du primaire s'approprient les ENT et quel est leur niveau d'acceptation de ce type d'outil. Pour y répondre nous avons réalisé deux études centrées sur l'utilisation de l'ENT ONE. La première avait pour objectif d'analyser les publications faites sur l'ENT afin de caractériser les différentes activités réalisées avec ONE. La seconde avait pour objectif d'identifier les facteurs de l'acceptation et/ou du refus de ce type de technologie.

Les résultats de la première étude ont mis en évidence 5 types de publications : communicationnelles internes, communicationnelles externes, pédagogiques individuelles, pédagogiques collaboratives et loisirs. Les publications communicationnelles sont prédominantes et consistent en des articles sur le blog et le cahier multimédia avec des photos et résumés des activités réalisées. Cela montre la volonté des enseignants du primaire d'associer les familles à la vie de l'école et donc la volonté « d'ouvrir l'école ». Les publications de type pédagogique arrivent en deuxième plan et sont présentes surtout au cycle 3 : CM1 et CM2. Elles corres-pondent à des leçons, textes rédigés par les élèves, exercices, vidéos éducatives, etc. L'étude a montré que l'appropriation était plutôt favorable car les utilisateurs avaient la possibilité d'imaginer de nouveaux usages et de les expérimenter. La facilité d'utilisation de ONE et son adéquation au public d'utilisateurs cibles sont les facteurs les plus explicatifs de ce constat.

La seconde étude a mis en évidence la présence des facteurs favorables à l'acceptation, en particulier la possibilité de travailler en collaboration et de communiquer facilement avec les familles. L'étude a aussi mis en évidence la présence de tensions liées à l'exercice et à la pratique du métier, notamment la surcharge de travail, l'augmentation des responsabilités professionnelles liées à l'extension de l'« espace-temps scolaire » et la faible implication des parents. Nous avons proposé différentes recommandations pour favoriser l'acceptation des ENT comme de produire un meilleur accompagnement au niveau local et une meilleure communication au niveau ministériel sur les responsabilités de chaque utilisateur de l'ENT.

Les résultats de la première étude convergent avec ceux issus de la seconde. Nous avons constaté, à travers l'analyse des publications et des entretiens, la volonté des enseignants d'ouvrir l'école aux parents d'élèves, de communiquer des informations concernant la vie de l'école sur l'ENT. Les publications dans les cycles 1 et 2 sont majoritairement de type communicationnel. Ces résultats sont intéressants car ils semblent contredire les études qui affirment une non-intégration de la part des enseignants des pratiques communicationnelles à destination des familles (IGEN, 2006), (Bruillard, 2011). Nous devons cependant noter que les contextes d'enseignement secondaire et primaire sont différents en ce qui concerne la relation avec les parents : dans l'enseignement primaire la relation avec les parents a une plus grande importance étant donné l'âge des enfants. La particularité de l'enseignant unique rend aussi cette relation plus facile et plus proche. Compte tenu de ces aspects de proximité avec les familles, l'acceptation de l'outil ONE se base beaucoup sur la communication école-familles.

En revanche, cette démarche d'ouverture ne semble pas être suivie par les parents. Les enseignants déplorent leur faible implication. Ceux-ci ne consulteraient pas du tout ou pas suffisamment les informations qui sont mises à leur disposition sur l'ENT. Notre étude ne propose malheureusement pas de recueil des opinions des parents, ni de bilan quantifié de leurs connexions ; cet axe serait intéressant dans le cadre d'une future recherche. Ces résultats permettraient d'affiner le point de vue des enseignants qui s'appuie pour l'instant uniquement sur les retours oraux des parents ou la visualisation de modifications qu'ils ont pu faire (comme l'ajout de photo sur le profil d'un élève de maternelle ou de CP). En réalité, beaucoup de parents peuvent consulter l'ENT sans laisser de « trace » visible, pouvant donner l'impression de non-implication. L'ajout d'indicateurs visibles de l'activité de consultation de l'ENT est donc fondamental pour montrer l'intérêt des parents et motiver les enseignants à poursuivre leurs usages. Ces fonctionnalités ont été intégrées dans la conception de ONE et vont être proposées. Un nouveau service intitulé « Statistiques » permettra aux enseignants de consulter la fréquentation de l'ENT par les autres utilisateurs (élèves, parents). Nous confronterons les observations factuelles d'usage des parents et leurs points de vue avec ceux des enseignants pour analyser plus finement cette question de l'engagement des parents lors d'une prochaine étude. Cela nous permettra en particulier de savoir si la perception des enseignants est fondée ou pas.

Les tensions relevées à propos des élèves et leur utilisation de l'ENT sont liées au manque d'infrastructure de l'école, à la question de la sécurité des élèves et à la question de la responsabilité du contenu publié par les élèves. Les tensions relevées ne semblent pas décourager les pratiques autour de l'ENT et donc l'appropriation de cet outil. La qualité de conception de la plateforme joue ici un rôle fort. On observe en effet dans les essais concrets d'appropriation de l'ENT, des actions menées afin de contrecarrer justement certaines tensions, comme la surcharge du travail et le manque d'infrastructure. Ces actions concernent l'encouragement du travail collaboratif qui peut permettre aux élèves de s'entraider et de se transmettre aussi les connaissances liées à l'utilisation de l'ENT ONE. Les projets collaboratifs permettent également de travailler à plusieurs sur un ordinateur et sont un moyen de pallier le manque d'infrastructure. Concernant la sécurité des élèves, les enseignants montrent qu'ils savent contourner cette problématique : ils créent des comptes élèves pour pouvoir contrôler les messages envoyés par les élèves ou font des cours spécifiques dédiés à la sensibilisation aux usages de l'ENT et d'Internet dans lequel ils essaient de responsabiliser les élèves vis-à-vis de leur utilisation. L'analyse des publications nous montre que les élèves s'approprient réellement l'ENT à partir du CM1. C'est le moment où ils deviennent utilisateurs actifs de l'ENT, capables de publier leur propre contenu. Un résultat intéressant est le fait que les élèves publient des contenus de tous types : péda-

**Elena CODREANU, Christine MICHEL, Marc-Eric BOBILLIER-CHAUMON, Olivier VIGNEAU**

gogiques (individuels ou collaboratifs), communicationnels (relatifs aux sorties scolaires ou activités internes) ou de loisirs. Ces publications ont un rôle « pédagogique » pour les enseignants dans la mesure où ils travaillent en même temps l'écriture, l'expression ou la grammaire. Elles les font « grandir » aussi puisqu'ils réalisent ainsi une partie des tâches communicationnelles dévolues précédemment aux enseignants. Elles les valorisent et les motivent puisque leurs rédactions, dessins ou poésies sont rendues accessibles à tous les parents d'élèves, à la classe et aux autres classes de l'école.

De manière générale, les enseignants sont les principaux initiateurs des usages. Ce sont eux qui distribuent les identifiants de connexion aux élèves et/ou parents, ce sont eux qui décident à qui ils vont ouvrir un service et à quelle activité servira tel ou tel blog ou cahier. Comme nous l'avons vu, l'appropriation consiste en un ensemble d'essais et expérimentations différentes qui mènent finalement à des usages pérennes et stables. On peut dire que c'est à travers les expérimentations répétées des activités sur l'ENT que se révèlent des situations critiques et des questionnements, des craintes et des tensions. Finalement, les enseignants peuvent expérimenter les services de l'ENT et sélectionner ceux qui conviennent au mieux à leurs exigences.

Cette étude comporte certaines limites. Le faible nombre d'utilisateurs peut poser des questions en termes de généralisation de nos résultats. Cet échantillon réduit (26 enseignants) est dû au fait que l'étude s'est déroulée pendant la phase d'expérimentation du projet ONE, phase qui a visé uniquement 12 écoles. Une ou deux classes ont généralement utilisé l'ENT par école. Notre étude est de plus assez circonscrite à une technologie spécifique au territoire français. Elle n'est pas forcément généralisable aux Virtual Learning Environments (VLE) proposés dans les pays anglo-saxons. Ces systèmes ont été construits principalement comme des outils pédagogiques, sur lesquels ont été ajoutés des modules administratifs (pour la gestion de notes et absences par exemple). En France, les ENT ont été conçus depuis le début pour répondre aux deux objectifs : de gestion (de la vie scolaire) et de pédagogie. Elle s'articule avec d'autres systèmes d'information nationaux[2] comportant en particulier des informations sur les utilisateurs (conformes aux règles de la CNIL) pour créer les comptes personnels, ou les classes des établissements pour créer des groupes automatiques dans l'ENT. Ces choix techniques présupposent une collaboration entre les écoles et l'État, ce dernier étant le prescripteur de l'ENT. Ces particularités contextuelles influencent et modulent l'acceptation et l'appropriation de l'outil par les utilisateurs. Par conséquent, les résultats de cette étude ne peuvent pas être généralisés à toutes les technologies éducationnelles, mais restent circonscrits au contexte précis des ENT français. C'est pourquoi notre approche se veut plutôt exploratoire et qualitative. Elle vise à décrire la construction de l'appropriation et de l'acceptation d'une technologie.

En conclusion, si l'acceptation et l'appropriation de l'ENT ONE sont plutôt positives c'est parce qu'il est bien conçu et plutôt bien adapté aux pratiques des enseignants. L'appropriation se fait de manière équivalente sur la base de communication avec les familles et par les activités de nature pédagogique. Notre étude est plutôt encourageante et elle évalue pour la première fois les activités concrètes faites autour d'un ENT d'écoles primaires. Les problèmes principaux restent néanmoins liés aux modalités de sa mise en œuvre. Les recommandations que nous formulons sont orientées vers le ministère et les directeurs d'école. Ces acteurs sont invités à apporter des précisions sur les limites de l'espace-temps scolaire et sur les règles de gouvernance et de communication les plus adaptées pour ces plateformes lorsque de très jeunes enfants, n'ayant pas de compétences en termes d'usages sociaux du numérique, sont concernés.

---

[1] Les animateurs TICE sont des enseignants qui ont pour mission le développement de l'utilisation de TICE (Technologies d'Information et de Communication pour l'Education) dans leurs circonscriptions. Ils sont sous la responsabilité des IEN et peuvent exercer à temps plein ou partiel.

[2] Le processus d'alimentation automatique de l'ENT est un processus défini par le Ministère de l'Education et correspond à la création des comptes ENT automatiques pour tous les futurs utilisateurs de l'ENT (enseignants, élèves, parents) d'un certain établissement. Cette alimentation se fait à partir des plateformes de gestion des enseignants (par exemples STS) et de gestion des élèves (Siècle -Système d'information pour les élèves en collèges et lycée et pour les établissements- dans les collèges et lycées et Base Elèves dans le primaire).

## BIBLIOGRAPHIE